\documentclass[10pt, conference]{IEEEtran}
\usepackage[english]{babel}
\usepackage[usenames]{color}
\usepackage{colortbl}
\usepackage{comment}
\usepackage{graphicx}
\usepackage{epsfig}
\usepackage{array, colortbl}
\usepackage{listings}
\usepackage{epstopdf}
\usepackage{multirow}
\usepackage{booktabs}
\usepackage{amsmath}
\usepackage{rotating}
\usepackage{subfig}
\usepackage{float}

\usepackage[obeyspaces,hyphens,spaces]{url}
\usepackage{balance}
\usepackage{fancybox}
\usepackage{scalefnt}
\usepackage[normalem]{ulem}
\usepackage{hyperref}

\makeatletter
\renewcommand{\paragraph}[1]{\noindent\textsf{#1}.}
\title{Log-based vs Graph-based Approaches to Fault Diagnosis}

\author{
    NGUYEN Mathis\\
    \emph{Dep. of Computer and Software Engineering, Polytechnique Montréal}\\
    \emph{Montréal, Canada}\\
    \emph{mathis.nguyen@polymtl.ca}\\[1ex]
    LAJNEF Mohamed Ali\\
    \emph{Dep. of Computer and Software Engineering, Polytechnique Montréal}\\
    \emph{Montréal, Canada}\\
    \emph{mohamed-ali.lajnef@polymtl.ca}
}

\begin{document}
\maketitle

\begin{abstract}
Modern distributed systems generate large volumes
of logs that can be analyzed to support essential AIOps tasks
such as fault diagnosis, which plays a crucial role in maintaining
system reliability. Most existing approaches rely on log-based
models that treat logs as linear sequences of events. However, such
representations discard the structural context between events that
are often present in execution logs, such as parent-child dependencies, fan-out (branching), or temporal features. To better capture these relationships, recent works on Graph Neural Networks (GNNs) suggest that representing logs as graphs offers a promising alternative. Building on these observations, this paper conducts a comparative study of log-based encoder architectures (e.g., BERT) and graph-based models (e.g., GNNs) for automated fault diagnosis. We evaluate our models on TraceBench \cite{tracebench}, a trace-oriented log dataset, and on BGL, a more traditional system log dataset, covering both anomaly detection and fault type classification. Our results show that graph-only models fail to outperform encoder baselines. However, integrating learned representations from log encoders into graph-based models achieves the strongest overall performance. These findings highlight conditions under which graph-augmented architectures can outperform traditional log-based approaches.
\end{abstract}

\section{Introduction}
\label{sec:introduction}

Fault diagnosis is a fundamental challenge in distributed systems, where failures can rapidly propagate across components and lead to performance degradation or service unavailability. To address this, automated learning-based methods for log analysis are now widely employed within the AIOps domain. While several approaches have been proposed, it remains unclear how different log representations influence diagnostic accuracy.

Traditional log-based encoders (e.g., BERT) process logs as linear sequences of events, capturing the semantic content of messages while being less sensitive to their structural execution patterns. In contrast, graph-based models (e.g., GNNs) can explicitly represent relationships between events, such as parent–child dependencies, execution hierarchies, or temporal ordering, thereby incorporating trace structure directly into the learning process. Comparing these two paradigms can therefore provide valuable insight into the respective contributions of event content and structural context to fault diagnosis.

This study focuses primarily on anomaly detection and extends the analysis to fault classification, a more challenging task that aims not only to detect failures but also to identify their underlying type.
We also investigate whether combining log-based encoders with graph-based models can leverage the strengths of both representations.

To guide this work, we address the following research questions:

\begin{itemize}
    \item \textbf{RQ1:} How effective are log-based encoders and graph-based models at distinguishing between normal and abnormal traces (\textit{anomaly detection})?\\
    This question examines how semantic-focused encoders (BERT) and structural-focused models (GNN) perform at detecting anomalies in execution traces. We compare the two pure modeling paradigms directly: encoder-only models that rely solely on event content and GNN-only models that rely solely on relational structure. This analysis clarifies the contribution of semantic information versus structural context to anomaly detection performance.

    \item \textbf{RQ2:} How effective are log-based encoders and graph-based models at classifying abnormal traces into their corresponding fault types (\textit{fault classification})?
    
    While most prior work focuses exclusively on anomaly detection, this question extends the analysis to fault classification, a more fine-grained diagnostic task. We perform the same semantic-versus-structural comparison as in RQ1, examining how each modeling paradigm supports the identification of fault types.

    \item \textbf{RQ3:} Does combining encoder-based representations with graph-based structural modeling improve fault diagnosis compared to using either independently?
    
    This question focuses on hybrid architectures that integrate semantic encoders with graph-based structural modeling. We assess whether combining event content and execution structure offers improvements over the two pure modeling paradigms across both anomaly detection and fault classification, all under the same experimental conditions.
\end{itemize}

\medskip
\noindent \textbf{Open Science}\ All code and experiments used in this study are available in an open-source repository: \url{https://github.com/mthsngn/Project-LOG6309E}.

\section{Related Work}
\label{sec:related-work}

Research on automated fault diagnosis in distributed systems has recently evolved along two main directions: log-based models \ref{log_based} that treat logs as sequences of events, and graph-based models \ref{graph_based} that exploit structural relations and execution context. Below, we review these two approaches and discuss how our work relates and differs.

\subsection{Log-based methods}
\label{log_based}

Traditional log analysis techniques treat system logs as linear sequences of events. Early methods extract templates from raw log messages and rely on rule-based or statistical heuristics (e.g., log frequency, regular expressions) to detect anomalies or classify faults. However, as system complexity and log volume increase, these approaches struggle to capture context-dependent, or previously unseen anomalies.

With advances in deep learning, modern log-based methods use neural network encoders such as LSTM-based models or Transformer architectures to learn semantic representations from logs treated as natural-language-like sequences. For example, \textbf{DeepLog} \cite{du2017deeplog} employs an LSTM to model normal event sequences and detect anomalies by flagging deviations from the predicted next event. \textbf{LogBERT} \cite{guo2021logbert} extends this idea with a BERT-based self-supervised framework that uses masked event prediction to capture richer contextual relationships between log messages. More recently, \textbf{LAnoBERT} \cite{lee2021lanobert} removes the need for a dedicated log parser by applying masked language modeling directly to raw log messages for unsupervised anomaly detection.

These log-based neural methods excel at capturing the semantic and contextual content of log messages. However, by treating logs primarily as sequences and not modeling their structural relationships, they may fail to capture anomaly-relevant signals embedded in the system’s execution flow. This limitation motivates the exploration of graph-based techniques (see Section~\ref{graph_based}).

\subsection{Graph-based Methods}
\label{graph_based}

Graph-based methods approach log analysis by organizing logs into execution traces or related event groups and modeling them as graphs, where nodes correspond to log events and edges encode structural dependencies.

\textbf{Logs2Graphs}~\cite{logs2graphs} converts the entire log corpus into a single attributed, directed, and weighted graph and applies a GNN-based detector to identify anomalies and highlight influential nodes.
\textbf{TraceGra}~\cite{tracegra} also adopts a structural perspective, representing microservice execution traces as graphs and using an encoder–decoder GNN to detect deviations in execution structure. Other techniques (e.g., \textbf{LogGD}~\cite{loggd}) generate a separate graph for each log sequence and use a Graph Transformer to detect anomalous patterns.
Beyond single-representation approaches, \textbf{DeepTraLog}~\cite{deeptralog} proposes a hybrid model that integrates log text with trace structure, showing that combining semantic and structural information can improve anomaly detection performance.

Overall, graph-based methods offer promising capabilities for leveraging execution structure, and research in this area remains active, with recent surveys \cite{graphsurvey} reviewing both their advantages and open challenges.

\subsection{Gaps and Positioning of Our Study}

Several research gaps remain in the existing literature.

Log-based neural encoders are highly effective at capturing the semantic and contextual content of log messages, but they generally ignore structural relations between events and may therefore struggle to exploit signals that arise from the system’s execution flow. Graph-based models address this limitation by explicitly encoding structural information such as parent–child links or temporal patterns, yet many faults do not manifest as structural anomalies and instead appear through subtle semantic differences in log messages \cite{graphsurvey}. 

Furthermore, transforming logs into execution graphs is often complex and dataset dependent, and GNNs are typically more expensive to train than encoder-based alternatives, raising questions about when their structural benefits genuinely justify this added complexity \cite{graphsurvey}.

Despite the diversity of existing approaches, there remains a lack of systematic empirical comparison between pure log-based methods and pure graph-based methods. Hybrid architectures that combine semantic encoders with structural modeling are relatively few, and existing proposals have not been directly compared against both pure paradigms \cite{deeptralog}. 

Finally, while anomaly detection has been widely studied \cite{ du2017deeplog,guo2021logbert,lee2021lanobert,logs2graphs,tracegra,loggd,deeptralog}, fault classification, a more fine-grained diagnostic task requiring the identification of the underlying fault type, has received far less attention. In particular, few studies jointly evaluate anomaly detection and fault classification in the context of hybrid architectures.

In light of these gaps, our work differentiates itself from prior studies in several ways:
\begin{itemize}
    \item We conduct a comparative evaluation of log-based encoder architectures (content) and graph-based models (structural) under a unified experimental setup.
    \item We examine whether hybrid models that combine log-based encoders with graph-based structural modeling can leverage the strengths of both representations and improve performance across tasks, assessed within the same setup.
    \item We extend the analysis beyond anomaly detection to include fault classification, a finer-grained and less explored diagnostic task.
\end{itemize}

\section{Approach}
\label{sec:approach}

\subsection{Overview}
\label{subsec:overview}

\begin{figure}[H]
    \centering
\includegraphics[width=1.0\linewidth,trim={0cm 22cm 0cm 0cm},clip]{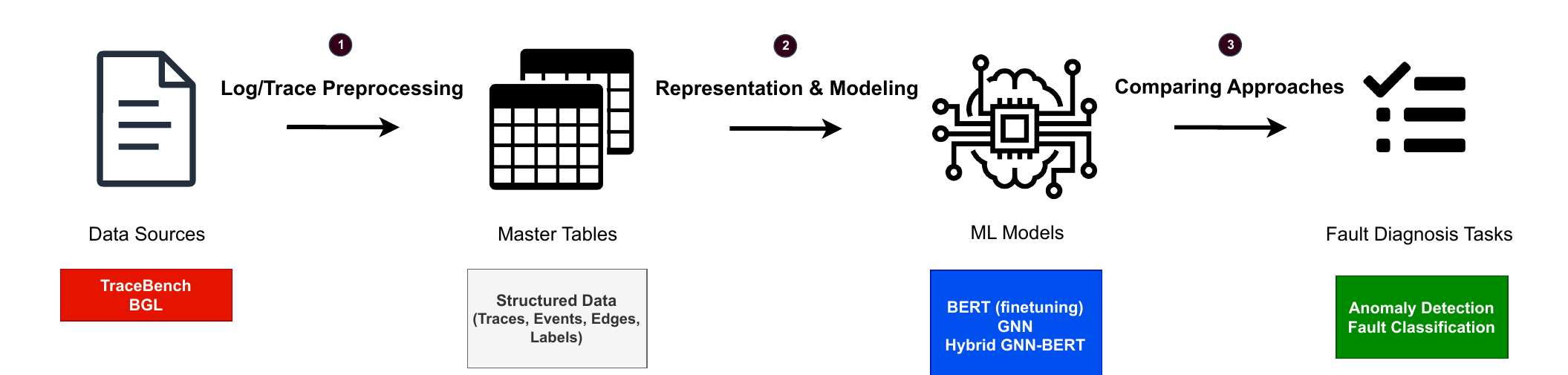}
    \caption{Overview of our end-to-end pipeline.}
    \label{fig:pipeline-overview}
\end{figure}

To address the research questions, we evaluate two complementary modeling directions for trace-based anomaly diagnosis: (1) sequential encoders based on Transformers, and (2) graph neural networks that leverage execution-level structure. We test both approaches on TraceBench \cite{tracebench}, and extend our experiments to one additional log dataset (BGL) to evaluate generalizability.
Figure~\ref{fig:pipeline-overview} presents an overview of the workflow used in this study.

The process consists of 3 main stages:

\begin{enumerate}
    \item \textbf{Log/Trace Preprocessing:}
    Log and trace data from the TraceBench and BGL datasets are preprocessed through several steps, including parsing, grouping events into execution traces, and labeling.
    This stage generates the \textit{master tables}, the common representation used by all models, which link traces to events, link events to one another through edges, associate each event with its content, as well as each trace with its labels.

    \item \textbf{Representation \& Modeling:}  
    From these structured tables, we derive three modeling paradigms:  
    a fine-tuned BERT model that learns semantic representations from event content;  
    a GNN model trained from scratch that captures structural information from a constructed event graph;  
    and a Hybrid GNN-BERT model that integrates both semantic embeddings and graph topology.
    
    \item \textbf{Comparing Approaches:}  
    We evaluate the 3 models under the same setup on the two fault-diagnosis tasks: Anomaly Detection \textit{(AD)} and Fault Classification \textit{(FC)}.
\end{enumerate}

\subsection{Datasets}

Our study was conducted on two datasets.

\begin{figure}[H]  
    \centering
    \includegraphics[width=0.45\textwidth]{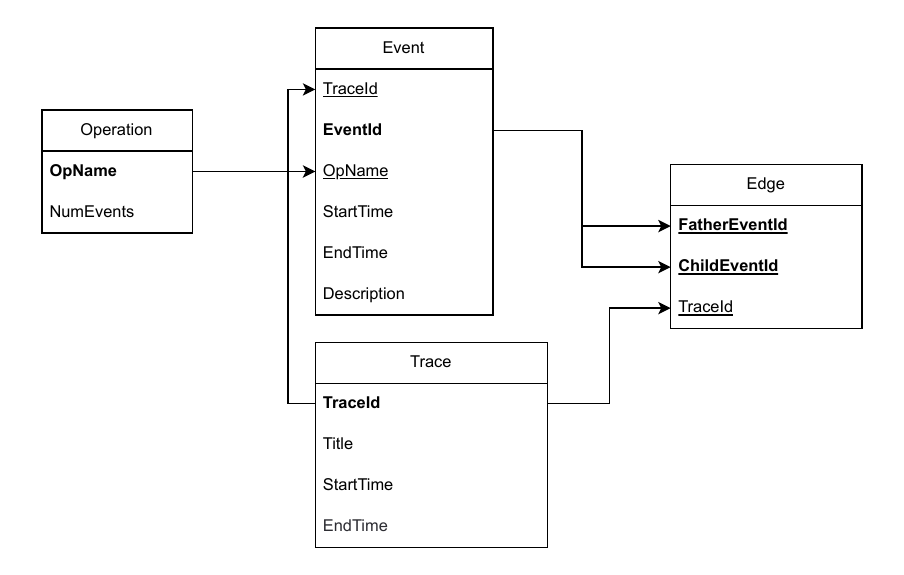}
    \caption{TraceBench Dataset (Simplified).}
    \label{fig:dataset-schema}
\end{figure}

The first, \textbf{TraceBench}, is an open dataset for trace-oriented monitoring, collected on an HDFS system deployed in a real IaaS environment. It contains approximately 370,000 execution traces, some of which include injected fault types introduced by the authors. Figure~\ref{fig:dataset-schema} illustrates the structure of the TraceBench dataset.

The dataset is organized into a set of relational tables. Each entry in the \textit{Trace} table corresponds to a complete execution instance of the system.
The \textit{Event} table records all events belonging to these executions, including both semantic attributes (e.g., operation name, description) and temporal metadata (start and end times).
The \textit{Edges} table specifies the parent–child relationships between events, defining the execution flow within each trace.

The injected faults to generate abnormal traces are as listed in Table~\ref{tab:fault-types}.

\begin{table}[H]
\centering
\scriptsize
\renewcommand{\arraystretch}{1.1}
\begin{tabular}{|l|l|p{5.7cm}|}
\hline
\textbf{Type} & \textbf{Fault} & \textbf{Description} \\
\hline
Process & killDN & Kill the HDFS processes on some datanodes \\ 
\cline{2-3}
& suspendDN & Suspend the HDFS processes on some datanodes \\
\hline
Network & disconnectDN & Disconnect some datanodes from network \\ 
\cline{2-3}
& slowHDFS & Slow all the HDFS nodes \\ 
\cline{2-3}
& slowDN & Slow some datanodes \\
\hline
Data & corruptBlk & Modify all the data blocks on some datanodes \\ 
\cline{2-3}
& corruptMeta & Modify all the metadata files on some datanodes \\ 
\cline{2-3}
& lossBlk & Delete all the data blocks on some datanodes \\ 
\cline{2-3}
& lossMeta & Delete all the metadata files on some datanodes \\ 
\cline{2-3}
& cutBlk & Remove some bits in all data blocks on some datanodes \\ 
\cline{2-3}
& cutMeta & Remove some bits in all metadata files on some datanodes \\
\hline
System & panicDN & Make the system panic on some datanodes \\ 
\cline{2-3}
& deadDN & Make the system dead on some datanodes \\ 
\cline{2-3}
& readOnlyDN & Make the system read-only on some datanodes \\

\hline
\end{tabular}
\caption{Default Fault Types in TraceBench.}
\label{tab:fault-types}
\end{table}

In contrast, the \textbf{BGL} dataset is a widely used benchmark for log-based anomaly detection originated from the IBM Blue Gene/L supercomputer at Lawrence Livermore National Laboratory. It consists of raw log files collected from large-scale distributed systems, each annotated to distinguish between normal and anomalous behaviors. It contains approximately 4.7 million log messages, each describing hardware or kernel-level activities such as parity errors, cache faults, or network issues. A typical normal entry looks like:
\begin{quote}
\texttt{1117838570 2005.06.03 R02-M1-N0-C:J12-U11 2005-06-03-15.42.50.363779 R02-M1-N0-C:J12-U11 RAS KERNEL INFO instruction cache parity error corrected}
\end{quote}
Each log line includes a timestamp, node identifier, component, severity level, and event description. These logs are labeled as either \textit{normal} or \textit{anomalous}, depending on whether they indicate system faults or expected operations.

\subsection{Log/Trace Preprocessing}

Our preprocessing pipeline differs significantly between the two datasets because they provide different levels of structure.

\medskip

\subsubsection{TraceBench}

TraceBench already provides well-structured traces. Each execution is recorded as a \textit{Trace} containing a sequence of \textit{Events}, and the \textit{Edge} table explicitly encodes parent–child relationships between events (Fig.~\ref{fig:dataset-schema}). 
As a result, preprocessing for this dataset consists of the following steps:

\textbf{Labeling.}  
Labels are derived from the structured naming pattern of each trace, which encodes the information needed for classification.
Using a simple regular-expression rule, we extract these labels to obtain the trace-level categories (normal or one of the injected fault types) and associate them with each trace entry.

\textbf{Filtering scenarios.}  
TraceBench includes traces collected under various cluster sizes or workload speeds.  
To ensure consistent experimental conditions, we restrict our study to the default collection scenario and discard all non-default configurations.
We also exclude the \textit{slowHDFS} fault type, which dominates the dataset and would otherwise distort the class distribution for Fault Classification.
After filtering, we retain a small but still sufficiently representative subset for our comparative study.

The resulting labeled and filtered tables constitute the initial master tables for our pipeline.

\medskip

\subsubsection{BGL}  Unlike TraceBench, BGL consists of raw, line-based system logs rather than trace-oriented records.
Therefore, it does not provide well-structured traces with event-edge relationships, and additional preprocessing is required to construct a trace-like structure compatible with our pipeline.

\textbf{Parsing.}  
Each log line is parsed to extract structured fields such as \texttt{Label}, \texttt{Timestamp} or, \texttt{EventTemplate}.  
We use the same regular expression as proposed in the paper~\cite{wu2024logrepresentation} to extract these fields consistently. 

\textbf{Grouping.}  
To approximate execution traces, parsed log lines are grouped into sessions.  
Following prior work~\cite{wu2024logrepresentation}, we divide the logs into fixed 6-hour windows, treating each window as a separate \textit{Trace} composed of its log messages.

\textbf{Labeling.}  
Each trace is labeled as \textit{normal} if all log entries contain the label “\texttt{-}”, or as \textit{abnormal} otherwise. Note that the labels are therefore only binary, as the dataset does not provide fault-type annotations.

\textbf{Constructing master tables.}  
Within each trace, individual log messages serve as \textit{Events} with their associated content (\texttt{EventTemplate}).  
Because BGL does not contain explicit causal relationships, we derive simple parent–child edges using timestamp ordering.
For consistency across datasets, we adopt the TraceBench naming conventions (e.g., \texttt{TraceId}, \texttt{EventId}, \texttt{FatherEventId}, etc.).

The resulting parsed, grouped, and labeled logs constitute the reconstructed master tables for our pipeline.

\medskip

Table~\ref{tab:session-stats} reports the number of traces in each dataset after preprocessing, together with their normal versus anomalous distribution. 

\begin{table}[H]
\centering
\caption{Trace counts and class distribution after preprocessing.}
\label{tab:session-stats}
\begin{tabular}{lrrr}
\hline
\textbf{Dataset} & \textbf{\# Traces} & \textbf{\# Normal} & \textbf{\# Abnormal} \\
\hline
TraceBench & 7{,}863 & 3{,}379 & 4{,}484 \\
BGL  & 824     & 411     & 413 \\
\hline
\end{tabular}
\end{table}

Table~\ref{tab:faulttype-stats} further breaks down the abnormal 
traces by their corresponding fault type for TraceBench specifically.

\begin{table}[H]
\centering
\caption{Fault type counts and distribution for TraceBench after preprocessing.}
\label{tab:faulttype-stats}
\begin{tabular}{l l r r}
\hline
\textbf{Dataset} & \textbf{Fault} & \textbf{\#} & \textbf{\%} \\
\hline
TraceBench & corruptBlk      & 345 & 4.39 \\
           & corruptMeta     & 354 & 4.50 \\
           & cutBlk          & 339 & 4.31 \\
           & cutMeta         & 355 & 4.51 \\
           & deadDN          & 367 & 4.67 \\
           & disconnectDN    & 400 & 5.09 \\
           & killDN          & 415 & 5.28 \\
           & lossBlk         & 333 & 4.24 \\
           & lossMeta        & 345 & 4.39 \\
           & panicDN         & 409 & 5.20 \\
           & readOnlyDN      & 66  & 0.84 \\
           & slowDN          & 365 & 4.64 \\
           & suspendDN       & 391 & 4.97 \\
\hline
           & \textbf{Total}  & \textbf{4,484} & \textbf{100.00} \\
\hline
\end{tabular}
\end{table}

\subsection{Representation \& Modeling}
\label{subsec:model&rep}

We investigate three complementary trace representations: a
log-based encoder model, a graph-based model, and a hybrid combination
of both. Each is detailed below.

\subsubsection{Encoder-based models }
Each trace is converted into a textual sequence of events. For every \texttt{TraceId}, events 
are ordered chronologically and concatenated, producing a single long content sequence for each trace. The exact formatting 
depends on the dataset:
{\small
\[
\text{Trace Encoding} =
\begin{cases}
\text{\texttt{OpName}:\texttt{Description} [SEP] …} & \text{(TraceBench)} \\
\text{\texttt{EventTemplate} [SEP] …}        & \text{(BGL)}
\end{cases}
\]

A simple modeling strategy is then to apply a standard sequence-classification pipeline: the text of each trace is first tokenized,
then padded or truncated to a fixed maximum length for batching, and finally processed by a
pretrained \textit{bert-base-uncased} Transformer model. The resulting
contextualised token embeddings are mean-pooled to obtain a trace-level
representation, which is fed to a lightweight MLP for either binary (normal
vs.\ abnormal) or multi-class (fault type) classification, and trained end-to-end by backpropagation. This approach was
used for the TraceBench dataset.

\medskip

A second strategy is preferred for datasets with highly variable and much longer
traces. Unlike TraceBench, where traces were relatively short and consistent,
BGL traces often contain thousands of events as a result of the
grouping stage, which makes simple truncation a serious limitation because most
of the sequence is never processed by the encoder.
To address this issue, we adopt a Multiple Instance Learning (MIL) inspired formulation. Instead of feeding the entire trace as a single sequence, each trace is partitioned into smaller fixed-size segments. Each segment is encoded independently by the Transformer \textit{bert-base-uncased}, and the classification head produces a score for every segment. These segment-level scores are then combined through a pooling operation (e.g., noisy-OR) to obtain a single trace-level prediction. This enables the model to exploit information from all parts of a long trace. This MIL-style strategy was applied to the BGL dataset.

\medskip

\subsubsection{Graph-based models}
Each trace is represented as a graph where
\[
    \text{Nodes} = \texttt{EventId}, \quad
    \text{Edges} = (\texttt{FatherEventId} \rightarrow \texttt{ChildEventId}).
\]
For each \texttt{TraceId}, we join the event and edge tables and construct directed edges from the father-child pairs in the edge table.

Node features include only structural and temporal attributes:
\begin{itemize}
    \item in-degree and out-degree in the trace graph,
    \item normalised position within the trace and normalised distance to the end,
    \item a normalised timestamp, using \texttt{StartTime} for TraceBench and
          \texttt{Timestamp} for BGL.
\end{itemize}

A two-layer GCN (\textit{GCNConv})
with ReLU activations processes the node features, and a global mean pooling layer
aggregates the node embeddings into a single trace-level representation. This trace embedding is then passed through a linear layer to produce the trace-level prediction for either binary (normal
vs.\ abnormal) or multi-class (fault type) classification.

\medskip

\subsubsection{Hybrid-models}

In addition to the two independent models, we introduce a hybrid architecture in which BERT-derived embeddings are injected into the GNN as additional semantic features. Instead of providing a single trace-level embedding, we compute a BERT representation for every event and concatenate this semantic vector with the structural and temporal node features used by the GNN. The graph model therefore operates on enriched node representations that capture both execution structure and log semantics. The resulting node features are processed by the same two-layer GCN as in the pure graph-based model, followed by global mean pooling and a linear classifier for the final trace-level prediction.

\subsection{Comparing Approaches}

\subsubsection{Training Setup}

All models are trained on the same data split. We use a stratified three-way split (70/15/15) to preserve the class distribution across training, validation, and test sets. The split is generated with a fixed random seed and the same procedure for every experiment.

Encoder-based models (BERT) are fine-tuned for 5 epochs with a batch size of 8, while graph-based models are trained for 30 epochs with a batch size of 16. Since the GNNs are trained from scratch whereas BERT is only fine-tuned, they require different numbers of epochs to reach convergence. Model selection is always performed using the best validation F1 score to ensure a fair comparison.

For anomaly detection, we use a binary cross-entropy objective, and for fault classification a multi-class cross-entropy loss. In both tasks, class imbalance is addressed through class-weighting, and all models are optimised using AdamW with a fixed learning rate.

\subsubsection{Evaluation methods}

The performance of the models is assessed using three commonly adopted metrics: \textit{Precision}, \textit{Recall}, and \textit{F1-score}. 
Model predictions are compared against the ground-truth labels to identify true positives (TP), false positives (FP), true negatives (TN), and false negatives (FN). 

\medskip

Using these values, the metrics are defined as:

\[
\text{Precision} = \frac{TP}{TP + FP}, \quad
\text{Recall} = \frac{TP}{TP + FN}, \quad
\]

\[
\text{F1} = \frac{2 \times \text{Precision} \times \text{Recall}}{\text{Precision} + \text{Recall}}
\]

For the multiclass fault-classification task, we report the
\textit{macro-averaged} scores, which compute each metric independently
for every class and then average them.

\subsubsection{Tasks per dataset}

TraceBench supports both anomaly detection (AD) and fault classification
(FC); BGL includes only anomaly labels and is therefore used only for anomlaly detection.

\section{Experimental Results}
\label{sec:results}

\subsection{RQ1. \textnormal{How effective are log-based encoders and graph-based models at distinguishing between normal and abnormal traces (\textit{anomaly detection})?}}

\medskip

\textbf{Motivation.}
Detecting anomalies is the first and most fundamental step in automated log analysis: before diagnosing a failure, a system must first recognize that its execution has deviated from normal behavior.
This research question aims to compare two widely used modeling paradigms for this task: encoder-based models that rely on event content, and the growing paradigm of graph-based GNN models that operate on execution structure. Through this analysis, we seek to clarify the relative importance and contributions of semantic information versus structural context for predicting anomalies, providing insights that can guide the design of future approaches for log-based anomaly detection.

\medskip

\textbf{Approach.}
To answer RQ1, we compare the encoder-only (BERT) model and the structure-only GNN model described in Section~\ref{subsec:model&rep}. Our pipeline, following the procedure detailed in Section~\ref{subsec:overview}, processes all datasets in a unified manner: after preprocessing, each trace is transformed into a token sequence for the BERT encoder and into a graph structure for the GNN. The two models are then trained in parallel on the same labels and evaluated under identical train/validation/test splits, ensuring a fair comparison. For completeness, it is worth noting that we use a standard sequence-classification formulation for TraceBench, whereas we handle BGL using a multiple-instance learning (MIL) scheme better suited to its very long trace sequences. Performance is assessed using precision, recall, and F1-score on both the TraceBench and BGL datasets.

\medskip

\textbf{Results.} 
Table~\ref{tab:ad_tracebench_bgl} summarizes the precision, recall, and F1-scores obtained by each model on TraceBench and BGL for anomaly detection, including a baseline Message Count Vector (MCV) combined with Logistic Regression (LR) approach, used as a reference point. Figures~\ref{fig:RQ1_TraceBench} and~\ref{fig:RQ1_BGL} provide a complementary visual comparison, helping illustrate the relative performance of semantic and structural models on each dataset for this task.

\begin{table}[h!]
\centering
\small
\caption{Precision, recall, and F1-score comparison on TraceBench and BGL (Anomaly Detection).}
\label{tab:ad_tracebench_bgl}
\setlength{\tabcolsep}{6pt}
\begin{tabular}{llccc}
\toprule
\textbf{Dataset} & \textbf{Model} & \textbf{Precision} & \textbf{Recall} & \textbf{F1-score} \\
\midrule
\multirow{3}{*}{TraceBench}
 & Baseline* & 0.826 & 0.892 & 0.858 \\
 & BERT                & \underline{\textbf{0.985}} & 0.887 & \underline{\textbf{0.934}} \\
 & GNN                 & 0.821 & \underline{\textbf{0.945}} & 0.870 \\
\midrule
\multirow{3}{*}{BGL}
 & Baseline* & \textbf{0.966} & \textbf{0.903} & \textbf{0.933} \\
 & BERT                & \underline{0.646} & \underline{0.855} & \underline{0.736} \\
 & GNN                 & 0.619 & 0.839 & 0.712 \\
\bottomrule
\multicolumn{5}{l}{\footnotesize *Baseline = Message Count Vector + Logistic Regression}\\
\multicolumn{5}{l}{\footnotesize \textbf{Bold} = best overall; \underline{Underline} = best between BERT and GNN.}
\end{tabular}
\end{table}

\begin{figure}[ht]
    \centering
    \includegraphics[width=0.95\linewidth]{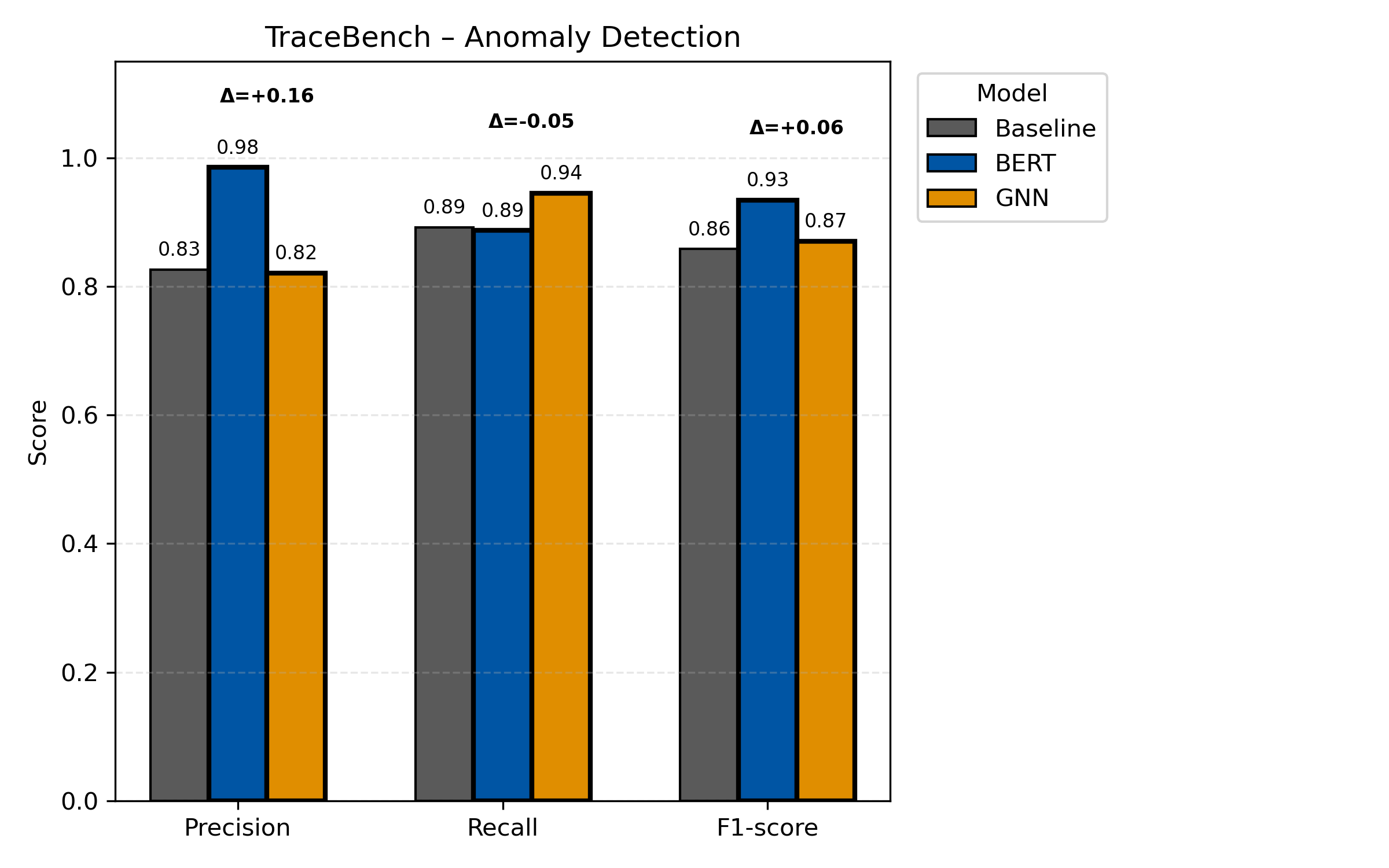}
    \caption{RQ1 – Comparison of semantic (BERT) and structural (GNN) models on TraceBench for anomaly detection.}
    \label{fig:RQ1_TraceBench}
\end{figure}

\begin{figure}[ht]
    \centering
    \includegraphics[width=0.95\linewidth]{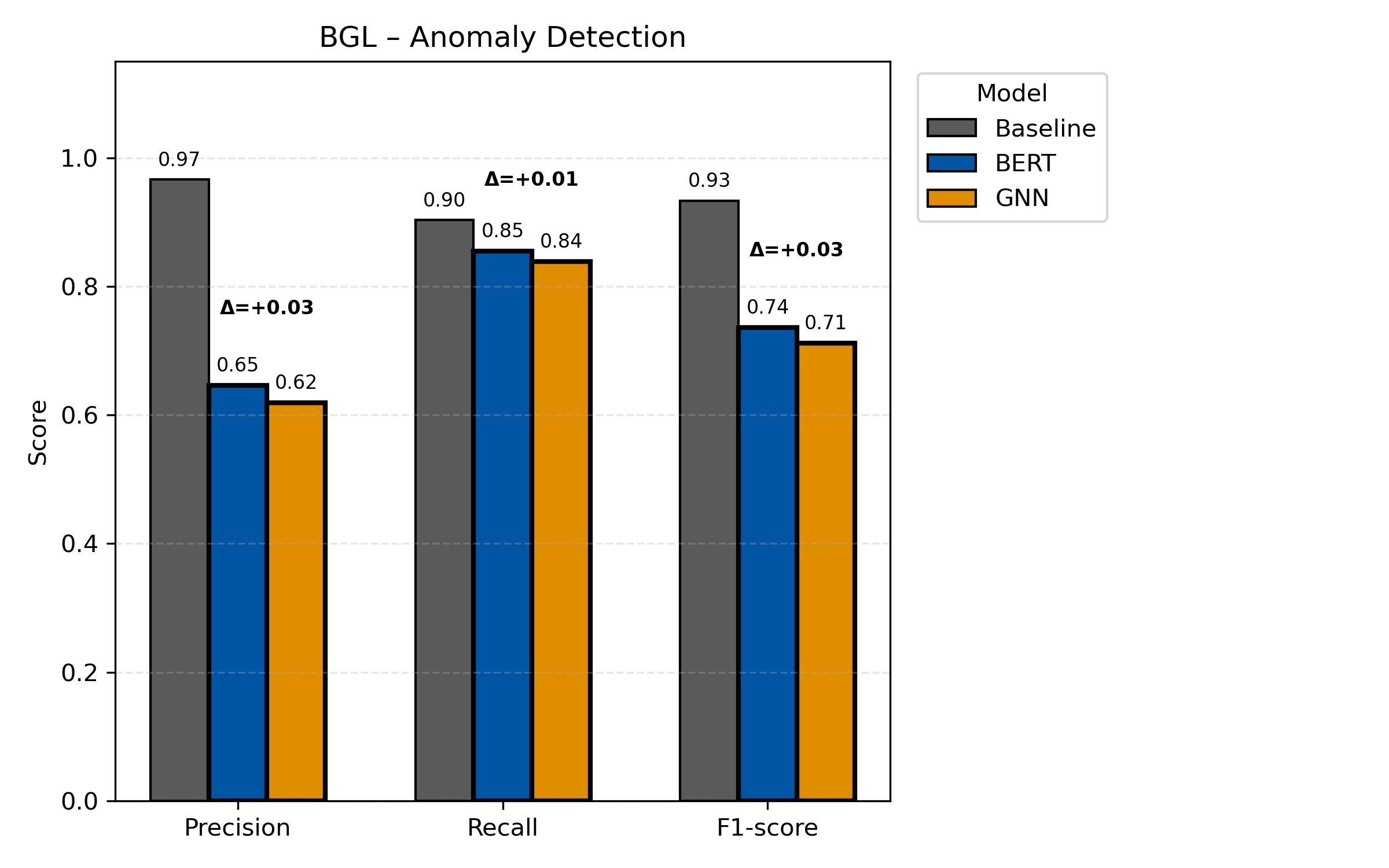}
    \caption{RQ1 – Comparison of semantic (BERT) and structural (GNN) models on BGL for anomaly detection.}
    \label{fig:RQ1_BGL}
\end{figure}

$\boldsymbol{\bullet}$\;\textbf{Log-encoder models achieve consistently better performance than graph-only models on anomaly detection.}

As shown in Table~\ref{tab:ad_tracebench_bgl}, on TraceBench, the BERT encoder obtains a higher F1-score (0.934) than the structure-only GNN (0.870). The gap is moderate, and the GNN even reaches the highest recall (0.945), meaning it successfully identifies most anomalous traces. When compared to the baseline (0.858), both BERT and the GNN improve performance, confirming the effectiveness of both semantic and structural representation-learning approaches on this dataset.

On BGL, both BERT and the GNN underperform relative to the baseline (0.933 F1). This behavior aligns with prior observations such as \cite{wu2024logrepresentation}, where simple count-based features combined with classical models (e.g., MCV + LR) achieved the strongest results among several tested approaches on this dataset. Several factors may contribute to this outcome. For BERT, the drop in performance may be explained by BGL’s very long traces with many events, which make semantic modeling at the trace level more challenging and require a MIL formulation instead of encoding the entire trace as a single sequence for training (Section~\ref{subsec:model&rep}). For the GNN, performance may similarly be affected by the fact that BGL is not inherently trace-oriented, meaning that our 6-hour grouping window produces synthetic traces that are not guaranteed to reflect actual execution flows. Even with these limitations, performance remains reasonable (with recalls of 0.855 for BERT and 0.839 for the GNN), and the relative ranking remains consistent with TraceBench: BERT stays slightly ahead of the GNN.

Overall, the results suggest that semantic information provides a notable superior contribution.

\medskip

$\boldsymbol{\bullet}$\;\textbf{Structural information alone is sufficient to achieve acceptable anomaly-detection performance.}

Despite ignoring all event content, the structure-only GNN achieves strong results, reaching 0.870 F1 on TraceBench and obtaining an F1-score of 0.712 on BGL, which remains relatively close to the encoder-only model ($\Delta = 0.03$). This shows that deviations in trace structure provide a strong enough signal to identify anomalous behavior, even without semantic information.

Overall, these results demonstrate that structural modeling alone can support reliable anomaly detection.

\subsection{RQ2. \textnormal{How effective are log-based encoders and graph-based models at classifying abnormal traces into their corresponding fault types (\textit{fault classification})?}}

\medskip

\textbf{Motivation.}
Fault classification is a more challenging and less studied task than anomaly detection, yet it is highly valuable because it provides the additional information needed to guide successful mitigation. Building on the motivation of RQ1, this research question examines how the same modeling paradigms perform when moving from binary anomaly detection to a fine-grained, multi-class prediction setting. By extending the analysis in this way, we assess whether the relative importance and effectiveness of semantic versus structural information shift at this more detailed level.

\medskip

\textbf{Approach.}
To answer RQ2, we apply the same modeling setup and follow the same pipeline as in RQ1, but now evaluate the models on the fault-type labels provided by the TraceBench dataset. Because fault classification is a multi-class task, performance is reported using macro-averaged precision, recall, and F1-score.

\medskip

\textbf{Results.} Table~\ref{tab:fc_tracebench} summarizes the precision, recall, and F1-scores obtained by each model on the TraceBench dataset for the multi-class fault classification task, including the baseline Message Count Vector (MCV) combined with multiclass Logistic Regression (LR). Figure~\ref{fig:RQ2} provides a graphical view of the same results.

\begin{table}[h!]
\centering
\small
\caption{Precision, recall, and F1-score comparison on TraceBench (Fault Classification).}
\label{tab:fc_tracebench}
\setlength{\tabcolsep}{6pt}
\begin{tabular}{llccc}
\toprule
\textbf{Dataset} & \textbf{Model} & \textbf{Precision} & \textbf{Recall} & \textbf{F1-score} \\
\midrule
\multirow{3}{*}{TraceBench}
 & Baseline* & \textbf{0.696} & 0.692 & \textbf{0.688} \\
 & BERT                & \underline{0.657} & \underline{\textbf{0.694}} & \underline{0.630} \\
 & GNN                 & 0.115 & 0.171 & 0.074 \\
\bottomrule
\multicolumn{5}{l}{\footnotesize *Baseline = Message Count Vector + Multiclass Logistic Regression}\\
\multicolumn{5}{l}{\footnotesize \textbf{Bold} = best overall; \underline{Underline} = best between BERT and GNN.}
\end{tabular}
\end{table}

\begin{figure}[ht]
    \centering
    \includegraphics[width=1.05\linewidth]{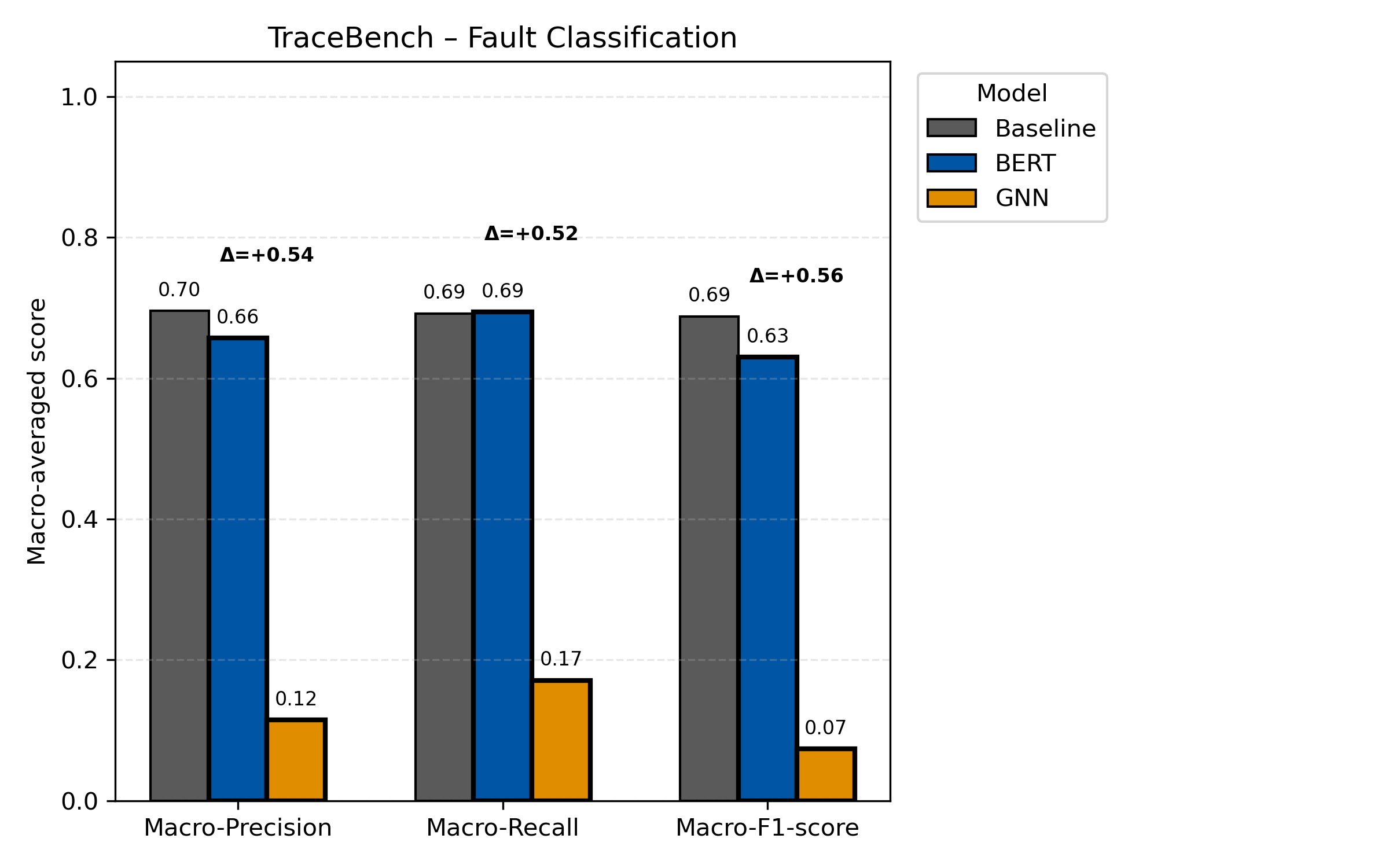}
    \caption{RQ2 – Comparison of semantic (BERT) and structural (GNN) models on fault classification.}
    \label{fig:RQ2}
\end{figure}

$\boldsymbol{\bullet}$\;\textbf{Log-encoder models outperform graph-only models on fault classification, as trace structure alone fails to discriminate fault types.}

The fault‐classification results in Figure~\ref{fig:RQ2} visually highlight a clear separation between semantic and structural models, with BERT showing performance gaps of more than $\Delta > 0.50$ over the GNN across all metrics. Table~\ref{tab:fc_tracebench} shows that the BERT encoder achieves substantially higher F1-performance than the structure-only GNN, reaching a score of 0.630 compared to only 0.074 for the GNN. Although BERT does not surpass the baseline MCV + multiclass LR (0.688 F1), it remains competitive and delivers the highest recall (0.694). 

In contrast, the GNN performs poorly across all metrics, with precision, recall, and F1-values far below both the baseline and the encoder. In fact, its accuracy is 0.107, which means it is only slightly above the $1/13 \approx 0.077$ chance level expected from random guessing given the 13 fault types. This points toward the idea that that structural information alone provides almost no discriminative signal for this task.

Overall, these results confirm that semantic information is essential for multiclass fault classification, while trace structure by itself seems to fails to provide enough signal that differentiates between the fault categories.

\subsection{RQ3. \textnormal{Does combining encoder-based representations with graph-based structural modeling improve fault diagnosis compared to using either independently?}}

\medskip

\textbf{Motivation.}
While RQ1 and RQ2 assess pure semantic and pure structural models independently, RQ3 investigates whether combining these two information sources leads to improved predictive performance. This question matters because hybrid architectures are increasingly adopted in practice. Demonstrating whether content and structure provide complementary signals directly informs whether hybrid modeling is worth the additional computational complexity.

\medskip

\textbf{Approach.}
To answer RQ3, we evaluate a hybrid BERT+GNN approach and compare it against the two pure paradigms considered in RQ1 and RQ2. As described in Section~\ref{subsec:model&rep}, we combine semantic and structural information within a single model by extending the structure-only GNN baseline: each event is first processed by the BERT encoder, and its embedding is then added as a node feature to the GNN. The hybrid model is trained on the same labels and evaluated using identical train/validation/test splits as the baselines. Evaluation covers both anomaly detection (TraceBench and BGL) and multi-class fault classification (TraceBench), using the metrics appropriate to each task.

\medskip

\textbf{Results.} Table~\ref{tab:all_models} reports the precision, recall, and F1-scores obtained by the baseline, the pure models (BERT and GNN), and the hybrid GNN+BERT model across both datasets and tasks. Figures~\ref{fig:RQ3-tracebench} and~\ref{fig:RQ3-bgl} provide complementary visualizations.

\begin{table}[h!]
\centering
\small
\caption{Precision, recall, and F1-score comparison on TraceBench and BGL (w/ Hybrid Models).}
\label{tab:all_models}
\setlength{\tabcolsep}{6pt}
\begin{tabular}{llccc}
\toprule
\textbf{Dataset} & \textbf{Model} & \textbf{Precision} & \textbf{Recall} & \textbf{F1-score} \\
\midrule
\multirow{4}{*}{TraceBench (AD)}
 & Baseline* & 0.826 & 0.892 & 0.858 \\
 & BERT                & 0.985 & 0.887 & 0.934 \\
 & GNN                 & 0.821 & 0.945 & 0.870 \\
 & Hybrid         & \textbf{0.986} & \textbf{0.969} & \textbf{0.978} \\
\midrule
\multirow{4}{*}{TraceBench (FC)}
 & Baseline* & 0.696 & 0.692 & 0.688 \\
 & BERT                & 0.657 & 0.694 & 0.630 \\
 & GNN                 & 0.115 & 0.171 & 0.074 \\
 & Hybrid          & \textbf{0.809} & \textbf{0.801} & \textbf{0.798} \\
\midrule
\multirow{4}{*}{BGL (AD)}
 & Baseline* & 0.966 & \textbf{0.903} & 0.933 \\
 & BERT                & 0.646 & 0.855 & 0.736 \\
 & GNN                 & 0.619 & 0.839 & 0.712 \\
 & Hybrid          & \textbf{0.982} & \textbf{0.903} & \textbf{0.941} \\
\bottomrule
\multicolumn{5}{l}{\footnotesize *Baseline = Message Count Vector + (Multiclass) Logistic Regression}\\
\multicolumn{5}{l}{\footnotesize Metrics for FC are macro-averaged.}\\
\multicolumn{5}{l}{\footnotesize \textbf{Bold} = best overall.}
\end{tabular}
\end{table}

\begin{figure}[ht]
    \centering
    \includegraphics[width=1.15\linewidth]{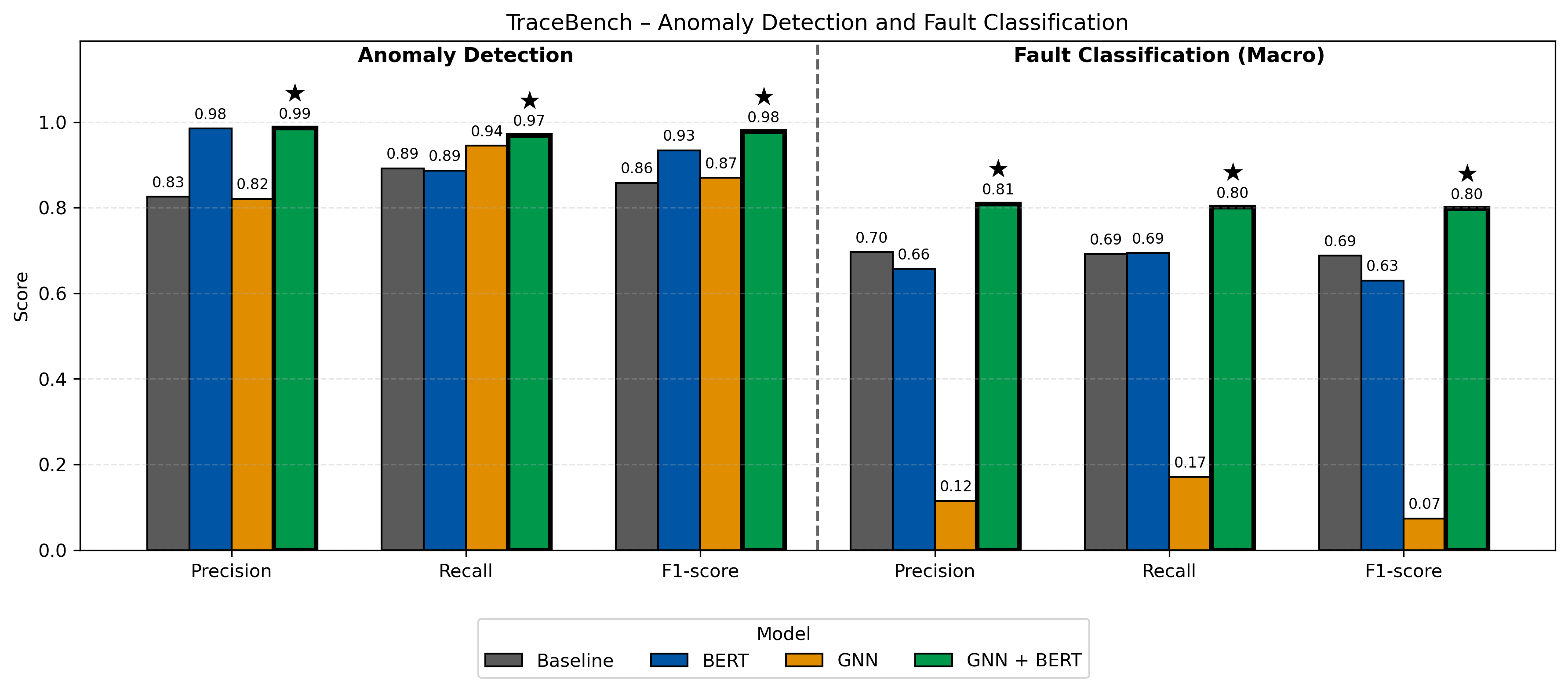}
    \caption{RQ3 – Hybrid model performance (GNN + BERT) on TraceBench for anomaly detection and fault classification.}
    \label{fig:RQ3-tracebench}
\end{figure}

\begin{figure}[ht]
    \centering
    \includegraphics[width=0.95\linewidth]{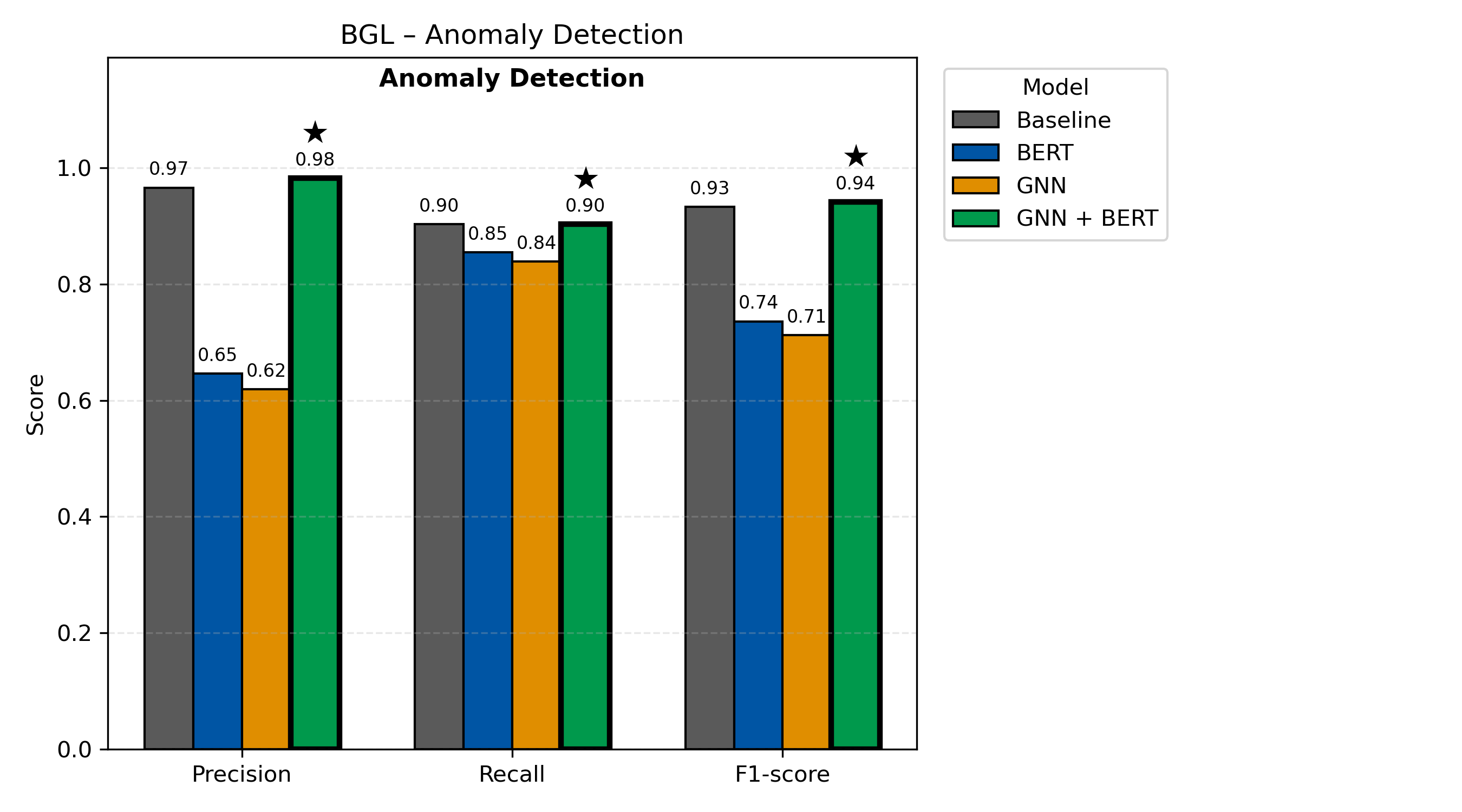}
    \caption{RQ3 – Hybrid model performance (GNN + BERT) on BGL for anomaly detection.}
    \label{fig:RQ3-bgl}
\end{figure}

$\boldsymbol{\bullet}$\textbf{Hybrid models achieve the best overall performance across all tasks and datasets.}

Table~\ref{tab:all_models} shows that the hybrid architecture systematically outperforms both pure semantic (BERT) and pure structural (GNN) models across all evaluated tasks.

When evaluating anomaly detection on TraceBench, the hybrid model reaches an F1-score of 0.978, substantially improving over both BERT (0.934), the GNN (0.870), and the baseline (0.858). It also attains the highest recall (0.969), suggesting that the hybrid approach can recover rare or subtle anomalous patterns that neither semantic nor structural information could capture on their own, despite the already strong performance of the individual models.

For fault classification on TraceBench, the hybrid approach delivers a major performance improvement: it reaches an F1-score of 0.798, far above BERT (0.630), the GNN (0.074), and the baseline (0.688). This indicates that enriching the GNN with semantic embeddings compensates for the inability of structural patterns alone to discriminate fault types while also surpassing purely semantic modeling, ultimately resulting in the best performance among all models.

Regarding anomaly detection on BGL, the hybrid model again achieves the best performance, with an F1-score of 0.941 compared to 0.736 for BERT and 0.712 for the GNN. It also surpasses the baseline (0.933) by a small margin. Despite the challenges posed by BGL’s long traces and the 6-hour window-based trace reconstruction, where the structure-only GNN showed limited effectiveness, the hybrid approach still succeeds in extracting a stronger signal, suggesting that some information becomes exploitable only when semantic and structural sources are combined.

Overall, these results demonstrate that merging semantic and structural information produces a consistently stronger model, with marked improvements across datasets and tasks, emphasizing the advantages of hybrid representations for fault diagnosis.

\section{Discussion}
\label{sec:disc}

Across the three research questions, several insights emerge about the respective roles of semantic and structural information in our trace-analysis setting.

First, both modeling paradigms provide meaningful signals for anomaly detection. As shown in our experiments, semantic patterns and structural deviations can each lead to strong performance, indicating that abnormal behavior can arise along multiple dimensions of the execution and can be effectively captured through different representational choices.

Second, BERT consistently delivers stronger performance across both anomaly detection and fault classification across our empirical results, suggesting that semantic content serves as the main informative signal. This aligns with the intuitive idea that semantic information provides the most discriminative indicators and captures the most relevant aspects of system behavior.

Next, the structure-only GNN demonstrates a distinct pattern: it performs well on anomaly detection but fails on fault classification. This indicates that structural deviations are informative enough to signal the presence of anomalous behavior, yet they do not provide sufficient discriminative power to distinguish between specific fault types. Still, the structural signal is not irrelevant; its contribution appears clearly in the performance of the hybrid approach, where incorporating structural patterns helps capture complementary aspects of system behavior that semantic encoders alone do not fully exploit, including for fault classification.

Finally, the hybrid architecture consistently achieves the best performance across all datasets and tasks. This “best of both worlds” behavior underscores the promise of hybrid representations and may also suggest that they can compensate for dataset-specific constraints, such as missing causal structure or noisy semantics, making them more robust overall.

\section{Threats to Validity}
\label{sec:thre-valid}

We have identified several threats that may impact the validity of our findings.

One limitation comes from how the models were configured. There are many possible encoder models based on content (e.g., BERT, RoBERTa, and many others), as well as many ways to design and configure them. Likewise, there are numerous ways to build a GNN, such as choosing different graph layers, aggregation methods, or incorporating additional information like edge features. All of these choices can influence performance. Exploring this entire space would make the comparison unfair, because some models might benefit more from tuning or added features than others. To avoid this, we followed simple and standard configurations and relied on widely used default hyperparameters across all models, while ensuring that the available information was still properly incorporated into the processing pipeline. This keeps the comparison more fair, but it also means that the full potential of each model may not be fully exploited.

The reliability of the results across repeated executions is also a concern, since they can differ depending on how the data is chosen and fed to the model, as well as other sources of randomness involved in training, including initialization, data shuffling, and nondeterministic GPU operations. To mitigate this, we used the exact same split for all models, ensuring that each approach was evaluated on identical data. However, statistical confidence could be further improved by repeating the experiments across multiple randomized splits and averaging the results, but this was not practical given our computational budget.

Another point concerns dataset considerations. While TraceBench provides logs with explicit trace structure and explicit edges, BGL contains only raw log lines, requiring heuristics such as timestamp ordering and hour-window grouping to reconstruct trace graphs. This type of data is less suited to trace-oriented analysis, though we still consider the experiment informative, as systems lacking structured tracing information or enriched metadata are far more typical in practice. To reduce the impact of these differences, we adopted a unified preprocessing pipeline to ensure consistent treatment across both data sources. However, the generalizability of our findings would benefit from additional systems representing a wider diversity of real-world logging patterns.

\section{Conclusion}
\label{sec:conclusion}

This study provides a systematic comparison of semantic, structural, and hybrid modeling approaches for trace-based anomaly detection and fault classification. Our results show that semantic information is the primary source of predictive accuracy, while structural patterns contribute substantially to anomaly detection and become particularly valuable when integrated with semantic features in a hybrid model. The findings also emphasize that model choice should reflect both the diagnostic task and the characteristics of the logging environment, ensuring that the selected representation can fully exploit the informative signals present in the data. These insights can support practitioners and researchers in designing more reliable log-analysis pipelines, both by guiding representation choices and by encouraging experimentation with hybrid architectures within their modeling strategy.

\balance
\bibliographystyle{IEEEtran}
\bibliography{project.bib}

@misc{tracebench,
  author       = {MTracer Team},
  title        = {TraceBench: A Benchmark Dataset for Trace-Oriented Monitoring},
  howpublished = {\url{https://mtracer.github.io/TraceBench/}},
  
  year         = {2021}
}

@article{logs2graphs,
   title={Graph Neural Networks based Log Anomaly Detection and Explanation}, 
      author={Zhong Li and Jiayang Shi and Matthijs van Leeuwen},
      year={2024},
      eprint={2307.00527},
      archivePrefix={arXiv},
      primaryClass={cs.SE},
      url={https://arxiv.org/abs/2307.00527}, 
}

@article{tracegra,
  author = {Chen, Jian and Liu, Fagui and Zhong, Guoxiang and Jiang, Jun and Xu, Dishi and Shi, Shangsong and Tan, Zhuanglun},
year = {2022},
month = {01},
pages = {},
title = {Tracegra: A Trace-Based Anomaly Detection for Microservice Using Graph Deep Learning},
journal = {SSRN Electronic Journal},
doi = {10.2139/ssrn.4211339}
}

@inproceedings{deeptralog,
  title={DeepTraLog: Unified Graph-Based Representation Learning for Microservice Anomaly Detection},
  author={Pengxin Cheng et al.},
  booktitle={ICSE},
  year={2022}
}

@article{loggd,
  title={LogGD:Detecting Anomalies from System Logs by Graph Neural Networks}, 
      author={Yongzheng Xie and Hongyu Zhang and Muhammad Ali Babar},
      year={2022},
      eprint={2209.07869},
      archivePrefix={arXiv},
      primaryClass={cs.SE},
      url={https://arxiv.org/abs/2209.07869}, 
}

@article{graphsurvey,
  title={A Survey of Graph-based Deep Learning for Anomaly Detection in Distributed Systems}, 
      author={Armin Danesh Pazho and Ghazal Alinezhad Noghre and Arnab A Purkayastha and Jagannadh Vempati and Otto Martin and Hamed Tabkhi},
      year={2023},
      eprint={2206.04149},
      archivePrefix={arXiv},
      primaryClass={cs.LG},
      url={https://arxiv.org/abs/2206.04149},
}

@inproceedings{du2017deeplog,
  title     = {DeepLog: Anomaly Detection and Diagnosis from System Logs through Deep Learning},
  author    = {Du, Min and Li, Feifei and Zheng, Guineng and Srikumar, Vivek},
  booktitle = {Proceedings of the 2017 ACM SIGSAC Conference on Computer and Communications Security (CCS)},
  pages     = {1285--1298},
  year      = {2017}
}

@article{guo2021logbert,
  title   = {LogBERT: Log Anomaly Detection via BERT},
  author  = {Guo, Haixuan and Yuan, Shuhan and Wu, Xintao},
  year    = {2021}
}

@article{lee2021lanobert,
  title   = {LAnoBERT: System Log Anomaly Detection based on BERT Masked Language Model},
  author  = {Lee, Yukyung and Kim, Jina and Kang, Pilsung},
  year    = {2021}
}

@article{wu2024logrepresentation,
  title={On the Effectiveness of Log Representation for Log-based Anomaly Detection},
  author={Wu, Xingfang and Li, Heng and Khomh, Foutse},
  journal={Empirical Software Engineering},
  year={2023},
  url={https://arxiv.org/abs/2308.08736}
}
\end{document}